\documentclass[prd,aps,showpacs,twocolumn,floats,nofootinbib]{revtex4}
\usepackage{epsfig}
\usepackage{graphicx,amsmath,amsfonts,amssymb,slashed,upgreek,color}
\definecolor{myblue}{rgb}{0,0.08,1} 
\definecolor{Dgreen}{rgb}{0.2,0.8,0.3}

\newcommand{\EE}{{\cal E}}

\newcommand{\KK}{{\cal K}}
\newcommand{\MM}{{\cal M}}
\newcommand{\OO}{{\cal O}}

\newcommand{\al}{\alpha}

\newcommand{\de}{\delta}

\newcommand{\ep}{\epsilon}
\newcommand{\ga}{\gamma}

\newcommand{\la}{\lambda}

\newcommand{\si}{\sigma}

\newcommand{\tr}{\mbox{tr}}

\newcommand{\be}{\begin{equation}}
\newcommand{\ee}{\end{equation}}

\newcommand{\lsim}{\stackrel{<}{\sim}}
\newcommand{\bea}{\begin{eqnarray}}
\newcommand{\eea}{\end{eqnarray}}
\newcommand{\bean}{\begin{eqnarray*}}
\newcommand{\eean}{\end{eqnarray*}}
\newcommand{\dd}{\partial}

\def\id{{\rm 1\kern -2.5pt I}}

\begin{document}

\title{Perturbations for massive gravity theories}

\author{Pietro Guarato and Ruth Durrer\\  }
\affiliation{D\'epartement de Physique Th\'eorique and Center for Astroparticle Physics, Universit\'e de 
Gen\`eve, 24 quai Ernest 
Ansermet,~CH--1211 Gen\`eve 4, Switzerland}
\email{pietro.guarato@unige.ch, ruth.durrer@unige.ch}

\date{\today}

\begin{abstract}
A theory of massive gravity depends on a nondynamical "reference metric" $f_{\mu\nu}$ which
is often taken to be the flat Minkowski metric.
In this paper we examine the theory of perturbations on a background with metric $\bar g_{\mu\nu}$ which 
does not coincide with the reference metric $f_{\mu\nu}$. We derive the mass term for general perturbations on this background and show that it generically is not of the form of the Fierz-Pauli mass term. We explicitly compute it for some cosmological situations and show that it generically leads to instabilities.
\end{abstract}


\maketitle

\section{Introduction}
\label{s:intro}
In recent years, interest in massive gravity theory has been rekindled. There are two main reasons for this: first, a graviton mass weakens gravity  on large scales and provides a natural mechanism of "degravitation" which can solve the cosmological constant problem~\cite{Dvali:2007kt,deRham:2007rw,deRham:2009rm}. If the graviton is massive, the range of gravity is finite and a cosmological constant does not gravitate. Second, if one fine tunes the graviton mass to $m_g \sim H_0$, where $H_0 \simeq 1.5\times 10^{-42}$GeV is the value of the Hubble constant, gravity weakens around this scale and such a modified gravity theory can explain the observed present accelerated  expansion of the Universe~\cite{Jha:2007ai,Sullivan:2011kv,Suzuki:2011hu,Hinshaw:2012fq,Planck:2013kta}; hence, it can play the role of dark 
energy~\cite{Koyama:2011yg,Gumrukcuoglu:2011zh,Langlois:2012hk,Gong:2012yv}.

In order to give the graviton, i.e., the degrees of freedom of the metric of spacetime a mass, one has to introduce
a reference metric in order to define a potential which gives energy to deviations away from the reference metric. For a scalar field or a vector field, this reference point is usually set to zero. For the metric this is not an option since the metric $f_{\mu\nu}=0$ is singular. 

There is also the possibility to avoid the reference metric but at the cost of nonlocal terms like for example $ m^2 \Box^{-1} G_{\mu\nu}$ in the equations of motion~\cite{Jaccard:2013gla}.
Such theories are usually not ghost free, but recently a solution where massive gravity can mimic dark energy for such a theory has been 
found~\cite{Maggiore:2013mea,Foffa:2013vma,Maggiore:2014sia}.

The most natural reference metric seems to be the Minkowski metric, $f_{\mu\nu}=\eta_{\mu\nu}={\rm diag}(-1,1,1,1)$, but in principle the reference metric is general~\cite{Hassan:2011tf}; also, other possibilities 
like a de Sitter reference metric~\cite{deRham:2012kf,Langlois:2012hk} have been considered.
Moreover, since time translation invariance is broken at very low energy, i.e. on cosmological scales, this might be an indication for a more general, less symmetric reference metric.

A generic quadratic term in the "metric perturbations" gives rise not only to three additional propagating gravitational modes which are necessary to complete the two massless modes to a massive spin-2 particle, but to an additional helicity-zero mode which is a ghost. To avoid this ghost, one has to introduce a mass term of a very specific form, the so-called Fierz-Pauli mass term~\cite{Pauli:1939xp}, but even in this case, as has been shown by Boulware and Deser~\cite{Boulware:1973my}, the ghost reappears at the nonlinear level.

Recently, de Rham, Gabadadze and Tolley (dRGT)~\cite{deRham:2010ik,deRham:2010kj} have proposed a nonlinear, polynomial generalization of the Fierz-Pauli mass term which is ghost free for an arbitrary reference metric $f$ and physical metric $g$. They have shown that the interactions between the different helicity modes can be at most fourth order in the Langrangian. The action is written in the form
\be
S = \frac{M_P^2}{2}\int \sqrt{-\det g}\left[R(g) - U(f,g)\right] \,,
\ee
where the second term, added to the usual Einstein-Hilbert action, takes into account the mass potential of the graviton.
This work has spurred a flurry of activity in massive gravity theories\footnote{Since 2010, 302 papers with "massive gravity" in the title have been submitted to the arXiv at the time of this writing.}. Especially, people want to investigate whether massive gravity can be at the origin of the observed accelerated cosmological expansion. For this, solutions which lead to an expansion history close to the one of the observable Universe have been studied~\cite{deRham:2012kf,Langlois:2012hk,Gong:2012yv}. 

To investigate cosmology in massive gravity, we of course cannot simply search for a background solution of massive gravity which reproduces the observed cosmological expansion history, but we also need to study perturbations on this cosmology which are relevant for the anisotropies of the cosmic microwave background and large-scale structure formation. This has been started for some specific cases e.g. in Refs.~\cite{Gumrukcuoglu:2011zh,DeFelice:2012mx,D'Amico:2012pi,Kuhnel:2012gh}.

This is also  where the present work sets in. We derive  the generic form of the graviton mass term in perturbation theory. For this we allow for an arbitrary reference metric $f_{\mu\nu}$ and a background solution $\bar{g}_{\mu\nu}$. We consider the true metric given by $g_{\mu\nu} = \bar{g}_{\mu\nu}+ h_{\mu\nu}$, where $h_{\mu\nu}$ is a small perturbation which we want to study up to quadratic order in the Lagangian. The first-order terms vanish due to the fact that $\bar g$ solves the equations of motion, and we are only interested in the second order. For the perturbed potential we can write up to second order in $h_{\mu\nu}$
\be\label{e:hMh}
\sqrt{-\det{g}} {U}(f,g) = \sqrt{-\det\bar{g}} \left[U(f,\bar g) +
  \MM^{\mu\nu\al\beta}h_{\mu\nu}h_{\al\beta}\right] \,.
\ee
The main goal of this work is to determine the tensor $\MM^{\mu\nu\al\beta}(f,\bar g)$ for arbitrary reference metric $f_{\mu\nu}$ and background $ \bar{g}_{\mu\nu}$.  We will find that for $f=\bar g$, the mass term is, as expected, the Fierz-Pauli combination. In this case, we know that also the higher order terms in ${h^\mu}_\nu$ are ghost free by construction. We show  that when $f\neq \bar g$ the quadratic mass term does not satisfy the Fierz-Pauli tuning.  However, this does not imply the presence of a ghost. In this nonperturbative case, it has to be checked that the constraint equations still project out the ghost. This has been done previously in Ref.~\cite{Hassan:2011tf}. However, it has also been shown recently that even the second scalar mode, which is "healthy" in vacuum, can become ghostlike in certain cases, e.g. in cosmology~\cite{Fasiello:2012rw}.

We finally discuss our mass term in a cosmological setting, where we also solve the perturbation equations for a special case.

The rest of the paper is organized as follows. In Sec.~\ref{s:metric} we derive the general form of $ \MM^{\al\beta\mu\nu}$. In Sec.~\ref{s:cos} we apply our result in cosmology and discuss it. In Sec.~\ref{s:con} we conclude. Some lengthy calculations are deferred to appendixes.

\noindent
{\bf Notation} We use the metric signature $(-,+,+,+)$. The reduced Planck mass $M_P$ is given by $M_P^2
=(8\pi G)^{-1}$, where $G$ denotes Netwon's gravitational constant. \\ Matrices are often denoted without indices, $g\equiv(g_{\mu\nu})$. In order to avoid confusion, determinants and traces are always clearly indicated as such, $\det g$ and $\tr \KK \equiv [\KK]$.

\section{Metric perturbations}
\label{s:metric}
Let us consider $\bar g_{\mu\nu}$ to be a solution to a given massive gravity theory with reference metric $f_{\mu\nu}$ and graviton potential
\be
U(f,g) 
    = - 2 m^2\left(U_2(\KK) + U_3(\KK) + U_4(\KK)\right) \label{e:pot} \\
\ee
 where
\be
{\KK^\mu}_\nu = \de^\mu_\nu  -(\sqrt{g^{-1}f})^\mu_\nu  ~~ \qquad \mbox{ and}
\ee

\bea
U_1(\KK) &=& [\KK] \, ,  \label{e:U1}\\
U_2(\KK) &=&  \frac{1}{2}\left([\KK]^2-[\KK^2]\right) \, ,\\
U_3(\KK) &=&  \frac{1}{6}\left([\KK]^3- 3[\KK][\KK^2] +2[\KK^3]\right)\,,\\
U_4(\KK) &=&  \frac{1}{24}\Big([\KK]^4-6[\KK]^2[\KK^2] +3[\KK^2]^2 + \nonumber\\
   && \hspace{0.36cm}  8[\KK][\KK^3] -6[\KK^4]\Big)  = \quad \det(\KK)\,.  \label{e:U4}
 \eea
Here we use the notation $[\KK] =\tr \KK = {\KK^\mu}_\mu$,  $[\KK^2] =\tr \KK^2 = {\KK^\mu}_\nu {\KK^\nu}_\mu$ and so forth. $U_1$ which does not appear in Eq.~(\ref{e:pot}) has been defined for later convenience. Notice that $\KK$ and therefore $U(f,g)$ vanish when $g=f$.

The square root of the matrix $g^{-1}f$ is just some matrix whose square is $g^{-1}f$. In general, this is not unique. However, if  $g^{-1}f$ is close to the identity, $g^{-1}f =\id + \ep$ with $|{\ep^\mu}_\nu| < 1/d$, where $d$ denotes the dimension of the matrix, we want to choose the root given by the convergent Taylor series,
\be
\sqrt{\id + \ep} = \id  +\sum_{k=0}^\infty\frac{(\frac{1}{2}-k)(\frac{1}{2}-k+1)\cdots \frac{1}{2}}{(k+1)!}\ep^{k+1}\,.
\ee

The potential $U(f,g)$ can be deformed by introducing arbitrary coefficients in front of $U_3$ and $U_4$,
\be\label{e:pot2}
U(f,g) = - 2 m^2\left(U_2(\KK) + c_3U_3(\KK)+c_4U_4(\KK)\right) \,.
\ee
In Ref.~\cite{Hassan:2011vm} it is shown that this is the most general potential for a ghost-free theory of massive gravity in four dimensions.

We now want to consider linear perturbations around a background solution with $\bar g_{\mu\nu} \neq f_{\mu\nu}$ for the massive gravity theory with potential~(\ref{e:pot2}). To derive the linear perturbation equations we  develop the Lagrangian
\be\label{e:lag}
L(g) = \frac{M_P^2}{2}\sqrt{-\det g}\Big(R(g) -U(f,g) \Big)
\ee
to second order in $h_{\mu\nu}$, the deviation of the true metric $g$ from the background, 
$g_{\mu\nu}= \bar g_{\mu\nu} + h_{\mu\nu}$. The kinetic term for $h_{\mu\nu}$ is determined by the 
Einstein operator, $\EE^{\mu\nu\al\beta}$, in curved spacetime~\cite{Hinterbichler:2011tt}
\bea \label{e:RgE}
\sqrt{\!-\det \!g}R(g)\! &=&\! \sqrt{\!-\det\bar g}\Big[R(\bar g)\!  - \!h_{\mu\nu}G^{\mu\nu}(\bar g) 
\nonumber\\ &&
\hspace{-1.5cm}+ h_{\mu\nu}\EE^{\mu\nu\al\beta}(\bar g)h_{\al\beta}   +\nabla_\mu V^\mu \Big]+\OO(h^3)\eea
with
\bea
\EE^{\mu\nu\al\beta}(\bar g) &=& -\frac{1}{2}\Big[\left(\bar g^{\mu\al}\bar g^{\nu\beta}-\bar g^{\mu\nu}\bar g^{\al\beta}\right)\Box   \nonumber \\  && \hspace{-1.5cm}+\left(\bar g^{\mu\nu}\bar g^{\al\rho}\bar g^{\beta\si}
+ \bar g^{\al\beta}\bar g^{\mu\rho}\bar g^{\nu\si} - \bar g^{\mu\beta}\bar g^{\nu\rho}\bar g^{\al\si} \right.  \nonumber \\ &&  \left. \hspace{-2.5cm} - \bar g^{\al\nu}\bar g^{\beta\rho}\bar g^{\mu\si}\right)\nabla_\rho\nabla_\si\Big]  +\frac{\bar R}{4}\left(\bar g^{\mu\al}\bar g^{\nu\beta} -\frac{1}{2}g^{\mu\nu}\bar g^{\al\beta}\right)\,.
\label{e:EE}
\eea
Here the covariant derivatives are taken with respect to the background metric $\bar g$ and $\Box =\bar g^{\rho\si}\nabla_\rho\nabla_\si$ is the d'Alembertian operator.
The kinetic term in square brackets in (\ref{e:EE}) is just the curved spacetime version of the well-known Einstein operator, see e.g.~\cite{deRham:2010ik} and the term proportional to $\bar R$ gives a contribution to the potential for $h_{\mu\nu}$ which vanishes in a flat background. 
This term looks like a mass term which does not satisfy the Fierz-Pauli tuning; however, this term is usually not harmful.
$G^{\mu\nu}$ in Eq.~(\ref{e:RgE}) is the Einstein tensor which solves the background equations of motion, and the total derivative $\nabla_\mu V^\mu$ is irrelevant for the equations of motion.

For $T^{\mu\nu} \neq 0$ there also comes a contribution to the mass term from the 
variation of the matter Langrangian which is of the form
\bea
\MM_{\rm mat}^{\mu\nu\al\beta} &=& \frac{1}{2}\frac{1}{\sqrt{-\det g}} \frac{\dd^2(\sqrt{-\det g} L_m)}{\dd g_{\mu\nu}\dd g_{\al\beta}}\bigg|_{g=\bar g} 
\nonumber \\  
  &=& \frac{1}{2}\frac{1}{\sqrt{-\det g}}
  \frac{\dd(\sqrt{-\det g} T^{\mu\nu}/2)}{\dd g_{\al\beta}}\bigg|_{g=\bar g} \,,
\eea
where $L_m$ denotes the matter Lagrangian. In the following we do not consider this model-dependent term. The result which we obtain is however strictly only valid in vacuum. This does not render it uninteresting as we expect that like the massless Einstein equations, also the massive equations have vacuum solutions where $\bar g$ differs widely from $f$ at least in certain regions of spacetime, like, e.g., the Schwarzschild solution. However, in a cosmological context, this matter-induced mass term does in principle also contribute.

We note in passing that the only difference of massive gravity theory to a bimetric theory of gravity is that our Lagrangian does not contain a kinetic term for the reference metric $f$. Massive gravity is therefore a theory with a 
"frozen-in" second metric $f$ which is not a dynamical element of the theory, but an "absolute spacetime". This is somewhat artificial. Actually, the beauty of general relativity where spacetime is dynamically determined by the matter content of the Universe is lost. Cosmological solutions for bimetric theories of gravity which  
add the term $ (M_P^2/2)\sqrt{-\det\! f}R(f)$ to the above Langrangian have also been studied~\cite{Volkov:2012cf,Volkov:2012zb,Koennig:2014aa}. 
 
The Einstein operator is symmetric under the exchange $(\mu\nu)\leftrightarrow (\al\beta)$. We could also symmetrize it in $\mu\nu$ and in $\al\beta$ but since we apply it only on the symmetric tensor $h_{\mu\nu}$ this does note make a difference. Furthermore, we omit the total derivative in Eq.~(\ref{e:RgE}) for simplicity.
 
We want  to determine the second-order perturbation of the potential. Up to second order in $h_{\mu\nu}$ the  potential is of the form
\bea
\sqrt{-\det g}U(f,g) &=& \sqrt{-\det\bar g}\Big[U(f,\bar g)  +
\MM^{\mu\nu}(f, \bar g)h_{\mu\nu}   \nonumber \\ && \hspace{0.61cm} + \MM^{\mu\nu\al\beta}(f,\bar g)h_{\mu\nu}h_{\al\beta}\Big] \,,
\eea
where
\bea \label{e:Mmunu}
\hspace{-2mm}\MM^{\mu\nu}(f,\bar g)\equiv \frac{1}{\sqrt{-\det g}}\frac{\partial
  (\sqrt{-\det g} U(f,g))}{\partial g_{\mu\nu}}\bigg|_{g=\bar{g}}, \\
\MM^{\mu\nu\al\beta}(f,\bar g)\equiv \frac{1}{2}\frac{1}{\sqrt{-\det g}}\frac{\partial^{2}
  (\sqrt{-\det g} U(f,g))}{\partial g_{\mu\nu}\partial g_{\al\beta}}\bigg|_{g=\bar{g}}. 
\eea

We consider perturbations around a solution $\bar g$ of the equations of motion.
 The terms linear in $h_{\mu\nu}$ in the Lagrangian therefore cancel due to the background equations of motion and we omit them in our discussion.

For noncommuting matrices $\sqrt{AB} \neq \sqrt{A}\sqrt{B}$, and we cannot simply expand 
$\sqrt{ g^{-1}f}= \sqrt{(\id + h)^{-1}\bar g^{-1}f}$ in $h=\left({h^\mu}_\al\right) = \left(\bar g^{\mu\nu}h_{\nu\al}\right)$.
Following~\cite{D'Amico:2012pi}, we therefore use the fact that the potential (\ref{e:pot2}) can also be written in the form
\bea
U(f,g) &=& - 2 m^2\left[a_0 + a_1U_1(\sqrt{ g^{-1}f})+ a_2U_2(\sqrt{ g^{-1}f}) 
       \right. \nonumber \\ &&  \left.  
  + a_3U_3(\sqrt{ g^{-1}f})\right] \,,  \label{e:pot3}
\eea
with 
\be
\begin{array}{ll}
a_0 = 6 +4c_3 + c_4, \quad  & a_1 = -(3+3c_3+c_4)\\
a_2 = 1 +2c_3 + c_4, \quad & a_3 =-c_3 - c_4 . 
\end{array}
\ee

Furthermore, as one can easily verify by bringing $\sqrt{ g^{-1}f}$ into triangular form,
\begin{subequations} \label{eq:t_i}
\begin{align}
& t_1 \equiv U_1(\sqrt{ g^{-1}f}) = \sum_i \la_i^{1/2} \, , \\
& t_2 \equiv U_2(\sqrt{ g^{-1}f}) = \sum_{i<k} \la_i^{1/2} \la_k^{1/2} \, , \\
& t_3 \equiv U_3(\sqrt{ g^{-1}f}) = \sum_{i<k<l} \la_i^{1/2} \la_k^{1/2} \la_l^{1/2} \, , \\
& t_4 \equiv U_4(\sqrt{ g^{-1}f}) = \sqrt{\la_1\la_2\la_3\la_4} \, ,
\end{align}
\end{subequations}
where $\la_i$ are the eigenvalues of $ g^{-1}f$, and $1\le i,k,l\le 4
$. Hence, we can write Eq.~(\ref{e:pot3}) as

\bea\label{e:pot4}
U(f,g) &=& - 2 m^2\left[a_0 + a_1t_1+ a_2t_2 + a_3t_3\right].  \label{e:pot3bis}
\eea

We define
\begin{subequations} \label{eq:s_i}
\begin{align}
& s_1 \equiv U_1({ g^{-1}f}) = \sum_i \la_i \, , \\
& s_2 \equiv U_2({ g^{-1}f}) = \sum_{i<j} \la_i \la_j \, , \\
& s_3 \equiv U_3({ g^{-1}f}) = \sum_{i<j<k} \la_i \la_j \la_k \, , \\
& s_4 \equiv U_4({ g^{-1}f}) = \la_1\la_2\la_3\la_4 \, .
\end{align}
\end{subequations}
We now use the following relations between the $t_j$ and  $s_i$ ($1\le j\le 3\, ,~~ 1\le i\le 4$):
\begin{subequations} \label{eq:relations}
\begin{align}
t_1^2 &= s_1 + 2 t_2 \, , \\
t_2^2 &= s_2 - 2 \sqrt{s_4} + 2 t_1 t_3 \, , \\
t_3^2 &= s_3 + 2 t_2 \sqrt{s_4} \, .
\end{align}
\end{subequations}
With this we can write the perturbations of $t_j$ in terms of perturbations of $s_i$ which in turn can be obtained from $g^{-1}f=(\id + h)^{-1}\bar g^{-1}f$.  We have to go to second order in the perturbations. The details of this lengthy calculation are given in Appendix~\ref{a:comp}, here we just present the result.
\begin{widetext}
\bea
\sqrt{-\det g}U(f,g) &=& \sqrt{-\det\bar g}\left[U(f,\bar g) + \MM^{\mu\nu\al\beta}(f,\bar g)h_{\mu\nu}h_{\al\beta} \right]  +\OO(h^3) \quad \mbox{with}\\
\MM^{\mu\nu\al\beta} &=& - m^2\left[a_0\MM_0^{\mu\nu\al\beta} +a_1\MM_1^{\mu\nu\al\beta} +a_2\MM_2^{\mu\nu\al\beta} +a_3\MM_3^{\mu\nu\al\beta} \right] \,, \label{e:MM}\\
\MM_0^{\mu\nu\al\beta} &=&\frac{1}{4}\bar g^{\mu\nu}\bar g^{\al\beta} -\frac{1}{4}\biggl(\bar g^{\mu\al}\bar g^{\nu\beta}+\bar g^{\mu\beta}\bar g^{\nu\al}\biggr) \\
\MM_j^{\mu\nu\al\beta} &=& \bar t_j\MM_0^{\mu\nu\al\beta} +\frac{1}{2}\left(\bar g^{\mu\nu} t_j^{\al\beta}+\bar g^{\al\beta} t_j^{\mu\nu}\right) + 2 t_j^{\mu\nu \al\beta}\,, \quad 1\le j\le 3 \,,\\
 t_j^{\mu\nu} &=& \left.\frac{\dd t_j}{\dd g_{\mu\nu}}\right|_{g=\bar g}\;, \qquad
 t_j^{\mu\nu\al\beta}~ = ~\left.\frac{1}{2}\frac{\dd^2 t_j}{\dd g_{\mu\nu}\dd g_{\al\beta}}\right|_{g=\bar g}\,.
\eea
\end{widetext}
Here $\MM_0^{\mu\nu\al\beta}$ is the second-order perturbation of the determinant $\sqrt{-g}$ and the quantities $ t_j^{\mu\nu}$ and $  t_j^{\mu\nu \al\beta}$ are
the first- and second-order derivatives of $t_j$ with respect to the metric components $g_{\mu\nu}$. Their full expressions are very cumbersome, they are given in Appendix~A.

Using the expressions given in  the Appendix, as a first check one can verify that this new quadratic potential for $h_{\mu\nu}$ reduces to the Fierz-Pauli mass term if $\bar g =f$,
\be
\MM^{\mu\nu\al\beta}(\bar g, \bar g) = - \frac{m^{2}}{4}\bigg[\bar g^{\mu\nu}\bar g^{\al\beta} -\frac{1}{2}\left (\bar g^{\mu\al}\bar g^{\nu\beta}+\bar g^{\mu\beta}\bar g^{\nu\al} \right) \bigg] \,,
\ee
where we have explicitly symmetrized with respect to the exchanges $ (\mu \leftrightarrow \nu)$, $(\al \leftrightarrow \beta)$.

Since the mass term given in Eq.~(\ref{e:MM}) is so complicated, it is very unlikely that it is of the Fierz-Pauli form in general. Nevertheless, as explained in the introduction, this does not mean that the theory has a ghost, when $\bar g \neq f$.

\section{Application to cosmology}
\label{s:cos}
\subsection{The mass term}
In this section we apply our finding in a cosmological setting.  To obtain a homogeneous and isotropic solution we first assume that both, $\bar g$ and $f$ are of the Friedmann-Lema\^{\i}tre  form with the same conformal time coordinate. To simplify the analysis we neglect curvature and set
\bea
\bar{g}_{\mu\nu}\mathrm{d}x^{\mu}\mathrm{d}x^{\nu}&=&a^{2}(t)(-\mathrm{d}t^{2}+\de_{ij}\mathrm{d}x^i\mathrm{d}x^j),\\
f_{\mu\nu}\mathrm{d}x^{\mu}\mathrm{d}x^{\nu}&=&b^{2}(t)(-\mathrm{d}t^{2}+\de_{ij}\mathrm{d}x^i\mathrm{d}x^j)\,.
\eea
Since the two metrics are proportional to each other, the mass term can only be of the form
\be\label{e:MMcos}
\MM^{\mu\nu\al\beta}(f,\bar g) = - m^{2}\biggl[\al\bar g^{\mu\nu}\bar g^{\al\beta} +\frac{\beta}{2}\left (\bar g^{\mu\al}\bar g^{\nu\beta}+\bar g^{\mu\beta}\bar g^{\nu\al} \right)\biggr]  \,.
\ee
In the cosmological situation $\al$ and $\beta$ depend only on time, but the expressions below in terms of  $r(t)=b(t)/a(t)$ are always correct when the two metrics $\bar g$ and $f$ are conformally related by $f =r^2\bar g$.

\begin{figure}[ht]
\includegraphics[width=7.5cm]{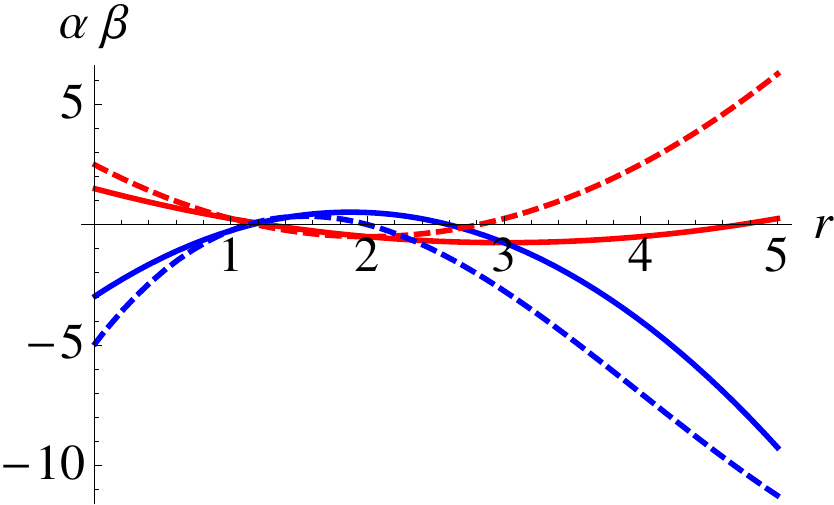}
\caption{The functions $\al(r)$ (red) and $\beta(r)$ (blue) are shown for two cases: $c_3=c_4=0$ (solid lines) and $c_3=1\,,~ c_4=0$ (dashed lines).}
\label{f:ab}
\end{figure}

Using the expressions in the Appendix and Eq.~(\ref{e:MM}), one obtains 
 \bea
\al(t) &=& \frac{1}{4}\bigg[1 +\left(1-r\right)\bigg\{\left(5-r\right) + 
  \notag \\   &&  \hspace*{-0.1cm}
c_{3}\left(4-2r\right) 
+c_{4}\left(1-r\right)\bigg\}\bigg]\,,   \label{e:al}\\
\beta(t) &=& -\frac{1}{4}\left[1 +\left(1-r\right)\left\{\left(11-4r\right) +
\right. \right.  \notag \\   &&  \hspace*{-0.1cm} \left. \left.
c_3\left(8-7r+r^{2}\right) +c_4\left(1-r\right)\left(2-r\right) \right\}\right]\,   \label{e:beta}.
 \eea
Evidently, for $r(t)=1$ or $a(t)=b(t)$ we recover the Fierz-Pauli mass term with $\al(t)= -\beta(t)=1/4$, for arbitrary values of $c_3$ and $c_4$, but since $r$ is time dependent, this value is not achieved in general. In Fig.~\ref{f:ab} we show the behavior of $\al$ and $\beta$ as functions of $r$ for some special values for $c_3$ and $c_4$.

In~\cite{Jaccard:2012ut}, it has been shown that on a fixed background the mass term~(\ref{e:MMcos}) for $\al\neq -\beta$ indicates the presence of a ghost with mass
\be
m^2_{\rm ghost} = \frac{(\al +4\beta)}{2(\al+\beta)}m^2 \,.
\ee
In our situation with $f\neq \bar g$ this is no longer true and the presence or absence of a ghost has to be investigated by other means. 
see e.g. Ref.~\cite{Hassan:2011tf}.

Let us contrast this result with the alternative possibility that $f$ and $g$ have the same {\em physical} time, which of course is not equivalent, 
\bea
\bar{g}_{\mu\nu}\mathrm{d}x^{\mu}\mathrm{d}x^{\nu}&=&-\mathrm{d}\tau^{2}+a^{2}(\tau)\de_{ij}\mathrm{d}x^i\mathrm{d}x^j, \label{e:g_phys} \\
f_{\mu\nu}\mathrm{d}x^{\mu}\mathrm{d}x^{\nu}&=&-\mathrm{d}\tau^{2}+b^{2}(\tau)\de_{ij}\mathrm{d}x^i\mathrm{d}x^j\,. \label{e:f_phys}
\eea
In this case the two metrics $f$ and $\bar g$ are no longer proportional and the mass term takes the more complicated form
\bea \label{e:MM_phys}
\MM^{0000} &=& - m^{2}\ga(\tau),  \\
\MM^{ij00} &=& - m^{2}\de(\tau)\bar g^{ij},  \\
\MM^{i0j0} &=& - m^{2}\ep(\tau)\bar g^{ij},  \\
\MM^{ijkl} &=& - m^{2}\bigg\{\rho(\tau)\bar{g}^{ij}\bar{g}^{kl}  + 
\notag \\   &&  \hspace*{-0.1cm} 
\frac{\sigma(\tau)}{2}\left[\bar{g}^{ik}\bar{g}^{jl}+\bar{g}^{il}\bar{g}^{jk}\right]\bigg\} \,.
\eea
Setting $r(\tau) = b(\tau)/a(\tau)$ we obtain
\bea
 \ga(\tau) &=& \frac{1}{4}\bigg[\left(1-r \right)\big\{(-6+3r)+ c_3 \left(-4+5r-r^{2}\right)+ 
\notag \\   &&  \hspace*{-0.1cm} 
 c_4 \left(-1+2r-r^{2}\right)\big\}\bigg], \\
  \de(\tau) &=& -\frac{1}{4}\bigg[1+(1-r)\big\{\left(5-r  \right)+ c_3 \left(4-2r\right)+ 
\notag \\   &&  \hspace*{-0.1cm}  
  c_4 \left(1-r\right)\big\}\bigg], \\
   \ep(\tau) &=& \frac{1}{4}\frac{1}{(1+r)}\bigg[1+\left(1-r \right)\big\{(5+2r-r^2) + 
\notag \\   &&  \hspace*{-0.1cm}   
   c_3 \left(4-r-r^{2}\right)+ 
 c_4 \left(1-r\right)\big\}\bigg], \\
     \rho(\tau) &=& \frac{1}{4}\bigg[1+\left(1-r \right)\big\{2+ c_3 \big\}\bigg], \\
\sigma(\tau) &=& -\frac{1}{4}\bigg[1+\left( 1-r \right)\big\{(5-r)  + c_3 \left(2-r\right)\big\}\bigg]. 
\eea

\begin{figure}[ht]
\includegraphics[width=8.5cm]{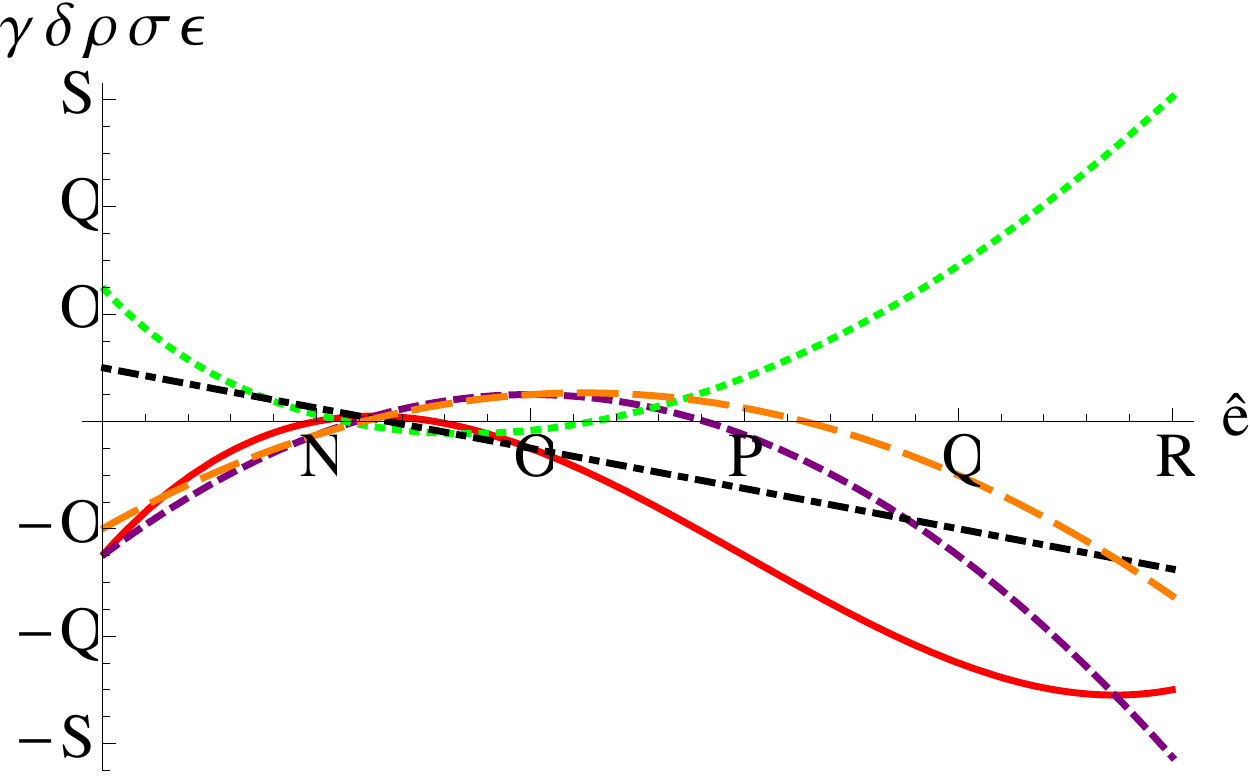}
\caption{The functions $\gamma(r)$ (red, solid line), $\delta(r)$
  (purple, dashed line), $\epsilon(r)$ (green, dotted line),
  $\rho(r)$ (black, dash-dotted line), and $\sigma(r)$ (orange,
  long-dashed line) are shown for the case $c_3=1$, $c_4=0$.}
\label{f:gammadelta}
\end{figure}

All other components of $\MM^{\mu\nu\al\beta}$ are determined by its symmetry under exchange $\mu\nu\leftrightarrow\al\beta$, $\mu\leftrightarrow\nu$ and $\al\leftrightarrow\beta$.
Again, when $r(\tau)=1$ or $a(\tau)=b(\tau)$, we reach the Fierz-Pauli tuning which corresponds to
$\ga =0$, $\rho =-\delta =-\sigma =1/4$, $\ep =1/8$. 
Note that in terms of the ratio $r$ $\de(r )=-\al(r )$ so that when writing $\MM^{ij00} = - m^{2}\phi(r )\bar g^{ij}\bar g^{00}$, we obtain the same expression for $\phi$ in both cases, equivalent physical time and equivalent conformal time. Interestingly, $c_4$ does not enter the expressions for $\rho$ and $\si$.
In Fig.~\ref{f:gammadelta} we show the behavior of $\delta$, $\gamma$, $\epsilon$, $\rho$, and $\sigma$ as functions of $r$ for the special case $c_3=1$, $c_4=0$. 

When $a(\tau)\neq b(\tau)$, the perturbations of these  solutions again violate the Fierz-Pauli tuning.

For a cosmological situation where the time directions of $f$ and $\bar g$ are boosted with respect to each other, the mass term is more complicated. However, this case would not allow for a homogeneous and isotropic solution and is therefore not relevant. The most general cosmological situation is $dt_f = r(t_g)dt_g$, where $t_f$ and $t_g$ denote the conformal times for the cosmological metrics $f$ and $\bar g$.

\subsection{Evolution of cosmological perturbations}

From Eq.~(\ref{e:lag}) we can derive the background equation of motion,
\be \label{e:back_eom}
\bar{G}_{\mu\nu}+\bar{\MM}_{\mu\nu}=M_P^{-2}\bar{T}_{\mu\nu},
\ee
where $\bar{G}_{\mu\nu}$ is the Einstein tensor for $\bar{g}_{\mu\nu}$, $\bar{T}_{\mu\nu}\equiv T_{\mu\nu}(\bar{g})$ and $\bar{\MM}_{\mu\nu}$ is the contribution from the mass term, which is calculated in Appendix~\ref{a:compMbar}.
For the cosmological form of the metrics (\ref{e:g_phys}) and (\ref{e:f_phys}) and the energy momentum tensor
\be
\bar{T}_{\mu\nu}=
\begin{pmatrix}
\bar{\rho} & 0 \\
0 & a^{2}\bar{p}\de_{ij} 
\end{pmatrix},
\ee
where $\bar{\rho}$ and $\bar{p}$ are the background energy density and pressure, respectively, we obtain the Friedmann equations 
\be\label{e:H}
\begin{split}
&3H^{2}+m^{2}\big[6-9r+3r^{2}+c_{3}(4-9r+6r^{2}-r^{3}) \\
&+ c_{4}(1-3r+3r^{2}-r^{3})\big]=M_P^{-2} \bar{\rho}
\end{split}
\ee
and
\be\label{e:Hdot}
2\dot{H}+3H^{2}+m^{2}\big[3-4r+r^{2}+c_{3}(1-2r+r^{2})\big]=-M_P^{-2}\bar{p},
\ee
where $H\equiv\dot{a}/a$ (the dot denotes the derivative with respect to physical time $\tau$).

We are interested in the question of whether perturbations of a cosmological solution have an instability due to the mass term, a ghost, in addition to the usual instability to gravitational clustering (Jeans instability). As is well known, the ghost always shows up in the scalar sector. Therefore, here we only analyze the scalar perturbation equations.  A more general analysis is presented in a forthcoming paper~\cite{next}.

The most general scalar perturbations  of the metric (in Fourier space) are of the form
\be
h_{\mu\nu}\equiv \delta g_{\mu\nu}=
\begin{pmatrix}
-2\phi & iak_{j}B \\
iak_{i}B & 2a^{2}(\psi\delta_{ij}-k_{i}k_{j}E)
\end{pmatrix} \,.
\ee
The perturbation equations resulting from this ansatz are Eqs.~(\ref{ea:00}), (\ref{ea:0inew}), (\ref{ea:ii}), and (\ref{eq:longit_traceless}), given in Appendix~\ref{a:eom}. These equations are still rather cumbersome, and a full analysis with cosmological expansion is given in~\cite{next}. Here we simply  analyze the presence of a ghost due to the mass term. For this, we simplify to the static solution $H\equiv0$ and matter domination $\bar p=0$. Inserting this in Eq.~(\ref{e:Hdot}), we find two possible solutions for $r$,
\be
r  =\left\{  \begin{array}{c}  1 \\ \frac{3+c_3}{1+c_3}=r_c\,\end{array}. \right.  
\ee
The first is simply Minkowski space with the Fierz-Pauli tuning. For this case, a brief analysis of the perturbation equations shows that there is no ghost but just one massive degree of freedom, namely $\psi$, as expected, the helicity 0 mode  of the massive graviton. For $r=r_c$, however, we obtain a static solution due to the presence of the mass term, which exists for $c_3\neq -1$. The positivity of the energy density $\bar{\rho}$ together with Eq.~(\ref{e:H}) then requires
$$P_1(c_3,c_4) =3 + 2 c_3 + 3 c_3^2-4c_4>0 \,.$$

 We can eliminate $\phi$ and $B$ using the constraint Eqs.~(\ref{ea:00}) and (\ref{ea:0inew}). We now consider the static case $r=r_c$ with vanishing matter perturbations $\de\rho=\de p=v-B=0$ since we want to study the evolution of the free gravitational field. Inserting $H=0$ and $r=r_c$ we obtain a system of the form
\be\label{e:osc}
\frac{d^2}{d\tau^2}\left(\begin{array}{c} \psi \\ \EE\end{array} \right) =  \left(m^2 A_0  + k^2  A_2 \right)
\left(\begin{array}{c} \psi \\ \EE\end{array} \right) \,,
\ee
where $\EE = m^2 E$. The matrices $A_0$ and $A_2$ are given by
\bea
A_0 &=&\left(\begin{array}{cc}  \frac{21+10c_3+9c_3^2-12c_4}{4(1 +  c_3)}  & 0 \\
Q(c_3) -\frac{4(1+c_3)(2+c_3)}{r_c P_1(c_3,c_4)} &r_c
\end{array}\right) \,,\\
A_2 &=&\left(\begin{array}{cc}  -\frac{1+c_3}{2}  &-\frac{P_1(c_3,c_4)}{4 (1 + c3)}\\
\frac{(1+c_3)(-5+c_3^2)}{r_c^2P_1(c_3,c_4)} &
\frac{-5+c_3^2}{2r_c(3+c_3)}\end{array}\right)   \,, \nonumber
\eea
where
\bean
Q(c_3) &=&\frac{33+27c_3-c_3^2-3c_3^3}{2(3+c_3)^2}\,.
   \eean
The eigenvalues of $A_0$ are
\bea
\la_{01} &=&\frac{3 + c_3}{1 + c_3} =r_c\\
\la_{02} &=& \frac{21 + 10 c_3 + 9 c_3^2 - 12c_4}{4(1 + c_3)} \,,
\eea
with eigenvectors
\bea
v_{01} &=& \left(\begin{array}{c} 0 \\ 1\end{array} \right) \\
v_{02} &=& \left(\begin{array}{c} \frac{3 P_1(c_3,c_4)}{
 4(1 + c_3)} \\ (A_0)_{21} \end{array} \right)\,. 
\eea
The fact that $\la_{01}>0$, indicates an exponential instability for small $k$.

\begin{figure}[ht]
\includegraphics[width=7.5cm]{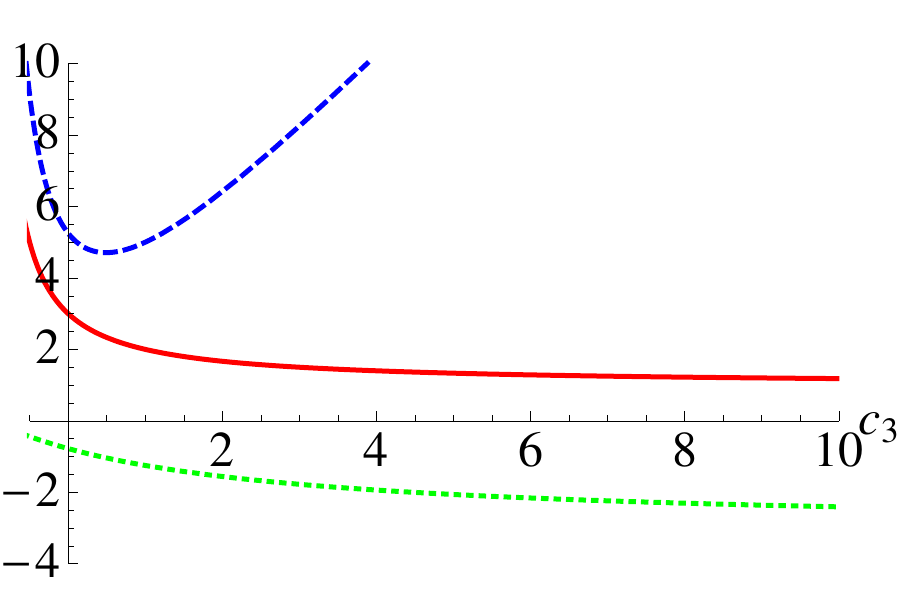}
\caption{The eigenvectors $\la_{01}$ (red, solid), $\la_{02}$ (blue, dashed) and $\la_{22}$ (green, dotted) are shown as functions of $-0.5<c_3<10$ for the case $c_4=0$.}
\label{f:evals}
\end{figure}

The eigenvalues of $A_2$ are
\bea
\la_{21} &=&0 \\
\la_{22} &=& -\frac{7+3c_3 }{r_c^2 (1+c_3)} \,,
\eea
with eigenvectors
\bea
v_{21} &=& \left(\begin{array}{c} -\frac{P_1(c_3,c_4)}{2(1+c_3)^2} \\ 1 \end{array} \right) \,, \\
v_{22} &=& \left(\begin{array}{c}-\frac{r_c^2 P_1(c_3,c_4)}{2(-5+c_3^2)} \\ 1 \end{array} \right)\,. 
\eea
The nonvanishing eigenvalues are shown as functions of $c_3$ for $c_4=0$ in Fig.~\ref{f:evals}. The situation for different values of $c_4$ is similar. Typically, one or both eigenvalues of $A_0$ are positive, which indicates an instability.

The eigenvalue $\la_{22}$ is negative for $c_3>-1$ so that high momentum modes are stable.
The value  $\la_{21}=0$ reflects the fact that in dRGT massive gravity, the second scalar mode does not really propagate~\cite{deRham:2010ik,deRham:2010kj}, but it also does not decouple as it does in the  Fierz-Pauli tuning. This comes from the choice of the potential $U(f,g)$. Nevertheless, as we have seen in this analysis, the mass term still leads to exponential instabilities as the eigenmodes of Eq.~(\ref{e:osc}) behave as $\exp(\pm\sqrt{\la_{0i}}mt)$ for small momenta.

At this point, it is not clear how the expansion of the Universe can mitigate this instability.
When the eigenvalue for the momentum, $\la_{22}$, is negative, there is still the chance that damping terms reduce the instability to a power law as long as $m^2\lsim H^2$. Hence it may be that the instability found here is not a disaster for the phenomenology of the
observable, expanding Universe. We study this issue in detail in a forthcoming publication~\cite{next}.

\section{Conclusions}
\label{s:con}
In this paper we have determined the form of the mass matrix
$\MM^{\mu\nu\al\beta}(f,\bar g)$ for fluctuations about some
background solution $\bar g$. We have shown that for $\bar g= f$ we
obtain the Fierz-Pauli mass term, whereas for $\bar g \neq f$ a more
general mass term is found. In the simple
case $f= r^2 \bar g$ the mass term is of the form
\be
\MM^{\mu\nu\al\beta}(f,\bar g) = - m^{2} \biggl[\al\bar g^{\mu\nu}\bar g^{\al\beta} +\frac{\beta}{2} (\bar g^{\mu\al}\bar g^{\nu\beta}+\bar g^{\mu\beta}\bar g^{\nu\al})\biggr] \,.
\ee 
We have calculated the functions $\al$ and $\beta$ in terms of $r$ and found that one recovers the Fierz-Pauli mass term only for $r=1$. Even if $r$ is a constant, $r=c\neq 1$, the mass term is different.

We have also calculated the mass term in the cosmological setting when $f$ and $\bar g$ have the same physical time but different conformal time. Also, in this case, when $\bar g \neq f$, the mass term differs from the Fierz-Pauli one.

We have briefly analyzed the consequence of this mass term in the case of "static cosmology" and have shown that even in this case, the mass term generically leads to instabilities. 

 In the future we want to study the contributions of matter, $T_{\mu\nu}$, to the mass term. This can be relevant in the cosmological cases studied here where matter can contribute significantly to the mass term. We plan to do this in a forthcoming paper~\cite{next}.  
 The main point of the present paper is the full calculation of the mass term for perturbations around an arbitrary background which can be used to study linear perturbation theory around arbitrary backgrounds and for an arbitrary reference metric.
 
\acknowledgments{ 
We thank  Claudia de Rham for important clarifications, and Lavinia Heisenberg, Michele Maggiore  and Mariele Motta for interesting discussions. 
This work is supported by the Swiss National Science Foundation.}
\vspace{0.9cm}

\appendix
\section{The computation of the perturbed potential}
\label{a:comp}
Here we present more details about the computation of $\MM^{\mu\nu\al\beta}(f,\bar g)$, and we give the detailed results.
With the help of Eq.~(\ref{eq:relations}) we can express the first- and second-order perturbations of $t_j$ in terms of those of $s_i$. Like for $t_j$ we set
\be
 s_i^{\mu\nu} = \left.\frac{\dd s_i}{\dd g_{\mu\nu}}\right|_{g=\bar g}\;,
\ee
\be
 s_i^{\mu\nu\al\beta} = \left.\frac{1}{2}\frac{\dd^2 s_i}{\dd g_{\mu\nu}\dd g_{\al\beta}}\right|_{g=\bar g}\,.
\ee 
To simplify the expressions we also introduce
\begin{equation}
A=2\sqrt{\bar{s}_4} (\bar{t}_1)^2 +2\bar{t}_3(-\bar{t}_1\bar{t}_2+\bar{t}_3),
\end{equation}
and the following combinations of first derivatives with respect to the background metric $\bar g$:
\begin{widetext}
\begin{subequations} \label{e:deltaBmunu}
\begin{align}
 B^{\mu\nu}_1 = \left(\bar{s}_4 \bar{t}_1-\sqrt{\bar{s}_4}
   \bar{t}_2 \bar{t}_3\right)  s^{\mu\nu} _1- \sqrt{\bar{s}_4} \bar{t}_3 s^{\mu\nu} _2-  \sqrt{\bar{s}_4} \bar{t}_1 s_3^{\mu\nu}+ \left(\bar{t}_3-\bar{t}_1 \bar{t}_2\right) s^{\mu\nu}_4 \, , \\
 B^{\mu\nu}_2=- \sqrt{\bar{s}_4}
   \bar{t}_3^2 s^{\mu\nu}_1- \sqrt{\bar{s}_4} \bar{t}_3 \bar{t}_1 s^{\mu\nu}_2- \sqrt{\bar{s}_4} \bar{t}_1^2  s^{\mu\nu}_3+ \left(\bar{t}_3-\bar{t}_1 \bar{t}_2\right) \bar{t}_1  s^{\mu\nu}_4 \, , \\
 B^{\mu\nu}_3=- \bar{s}_4 \bar{t}_3  s^{\mu\nu}_1- \bar{s}_4
\bar{t}_1 s^{\mu\nu}_2+ \sqrt{\bar{s}_4} \left(\bar{t}_3 -\bar{t}_1
   \bar{t}_2\right) s^{\mu\nu}_3  + \left(\sqrt{\bar{s}_4} \bar{t}_1+\bar{t}_2 \left(\bar{t}_3-\bar{t}_1 \bar{t}_2\right)\right) s^{\mu\nu}_4 ,
\end{align}
\end{subequations}
With this the first derivatives of the $t_j$ can be written as
\be \label{e:deltatmunu}
 t_j^{\mu\nu} = \frac{1}{A\sqrt{\bar{s}_4}} B_j^{\mu\nu} \, . 
\ee
To obtain the second derivatives we have to derive Eq.~(\ref{eq:relations}) a second time.
A rather cumbersome but straightforward calculation leads finally to
\begin{subequations} \label{eq:delta2t}
\begin{align}
 t^{\mu\nu\al\beta}_1 = & \left(\frac{ \bar{t}_2 \bar{t}_3}{A^3 \bar{s}_4}-\frac{ \bar{t}_1}{A^3 \sqrt{\bar{s}_4}}\right) B^{\mu\nu}_1 B^{\al\beta}_1 
-\frac{ \bar{t}_3}{A^3 \bar{s}_4} \biggl( 
B^{\mu\nu}_1 B^{\al\beta}_3 + 
B^{\al\beta}_1 B^{\mu\nu}_3  - B^{\mu\nu}_2 B^{\al\beta}_2 \biggr) 
+ \frac{ \bar{t}_1}{A^3 \bar{s}_4} B^{\mu\nu}_3 B^{\al\beta}_3 \notag \\ 
   & - \frac{ \bar{t}_1}{2 A^2 \bar{s}_4}\biggl( 
B^{\mu\nu}_2 s^{\al\beta}_4 +
B^{\al\beta}_2 s^{\mu\nu}_4  \biggr) 
   +\left( \frac{ \bar{t}_1 \bar{t}_2}{4 A
   \bar{s}_4^{3/2}}-\frac{\bar{t}_3}{4 A \bar{s}_4^{3/2}}\right) s^{\mu\nu}_4 s^{\al\beta}_4   \notag \\ 
   &+\left(\frac{ \sqrt{\bar{s}_4} \bar{t}_1}{A}-\frac{\bar{t}_2 \bar{t}_3}{A} \right) s^{\mu\nu\al\beta}_1 -
   \frac{\bar{t}_3}{A} s^{\mu\nu\al\beta}_2 -\frac{ \bar{t}_1}{A} s^{\mu\nu\al\beta}_3 +\left(\frac{ \bar{t}_3}{A \sqrt{\bar{s}_4}} - \frac{ \bar{t}_1
   \bar{t}_2}{A \sqrt{\bar{s}_4}}\right) s^{\mu\nu\al\beta}_4   
   \, , \\
 t^{\mu\nu\al\beta}_2 = & ~ \frac{ \bar{t}_3^2}{A^3 \bar{s}_4} B^{\mu\nu}_1B^{\al\beta}_1-\frac{\bar{t}_1 \bar{t}_3}{A^3 \bar{s}_4}\biggl(
B^{\mu\nu}_1 B^{\al\beta}_3 +
B^{\al\beta}_1 B^{\mu\nu}_3 -B^{\mu\nu}_2 B^{\al\beta}_2 \biggr)+\frac{ \bar{t}_1^2}{A^3 \bar{s}_4} B^{\mu\nu}_3 B^{\al\beta}_3 \notag \\  
   &-
\frac{\bar{t}_1^2}{2 A^2 \bar{s}_4}\biggl( 
B^{\mu\nu}_2 s^{\al\beta}_4 +
B^{\al\beta}_2s^{\mu\nu}_4  \biggr) +\left(\frac{\bar{t}_1^2 \bar{t}_2}{4 A \bar{s}_4^{3/2}}-\frac{\bar{t}_1 \bar{t}_3}{4 A
   \bar{s}_4^{3/2}}\right)  s_4^{\mu\nu} s^{\al\beta}_4
\notag \\  
   &   -\frac{ \bar{t}_3^2}{A}s^{\mu\nu\al\beta}_1-\frac{ \bar{t}_1
   \bar{t}_3}{A}s^{\mu\nu\al\beta}_2 - \frac{ \bar{t}_1^2}{A}s^{\mu\nu\al\beta}_3 + \left(\frac{ \bar{t}_1 \bar{t}_3}{A
   \sqrt{\bar{s}_4}}  - \frac{ \bar{t}_1^2 \bar{t}_2}{A \sqrt{\bar{s}_4}} \right)s^{\mu\nu\al\beta}_4 \, , \\
 t^{\mu\nu\al\beta}_3 = & \frac{\bar{t}_3}{A^3 \sqrt{\bar{s}_4}}B^{\mu\nu}_1 B^{\al\beta}_1-\frac{\bar{t}_1}{A^3 \sqrt{\bar{s}_4}} \biggl(
B^{\mu\nu}_1B^{\al\beta}_3 + 
B^{\al\beta}_1B^{\mu\nu}_3  -B^{\mu\nu}_2 B^{\al\beta}_2\biggr)+
\left(\frac{\bar{t}_1 \bar{t}_2}{A^3 \bar{s}_4}-\frac{\bar{t}_3}{A^3\bar{s}_4}\right) 
B^{\mu\nu}_3 B^{\al\beta}_3 \notag \\   
    &+\left(\frac{ \bar{t}_3}{2 A^2\bar{s}_4} - \frac{ \bar{t}_1
        \bar{t}_2}{2 A^2 \bar{s}_4} \right)\biggl( 
B^{\mu\nu}_2s^{\al\beta}_4 +
B^{\al\beta}_2s^{\mu\nu}_4  \biggr)
    + \left(\frac{\bar{t}_1 \bar{t}_2^2}{4 A \bar{s}_4^{3/2}}-\frac{\bar{t}_1}{4 A
   \bar{s}_4}-\frac{ \bar{t}_2 \bar{t}_3}{4 A \bar{s}_4^{3/2}} \right)s^{\mu\nu}_4 s^{\al\beta}_4   \notag \\
   &  -\frac {\sqrt{\bar{s}_4} \bar{t}_3}{A}
   s^{\mu\nu\al\beta}_1 -\frac{\sqrt{\bar{s}_4}\bar{t}_1}{A}s^{\mu\nu\al\beta}_2  +\left(\frac{ \bar{t}_3}{A} -\frac{ \bar{t}_1
       \bar{t}_2}{A}\right)s^{\mu\nu\al\beta}_3 + \left( \frac{\bar{t}_1}{A} +\frac{\bar{t}_2 \bar{t}_3}{A \sqrt{\bar{s}_4}} -\frac{ \bar{t}_1 \bar{t}_2^2}{A \sqrt{\bar{s}_4}}\right)s^{\mu\nu\al\beta}_4 \,.
  \end{align}     
   \end{subequations}
With this we have expressed the derivatives of the quantities $t_j$ in terms of those of the $s_i$, but the latter can be obtained directly by expanding the matrix 
$$g^{-1}f = \left(\bar{g}(1+h)\right)^{-1}f \approx (1-h +h^2)\bar g^{-1}f \,. $$
Here $h$ denotes $({h^\mu}_\nu) = (\bar g^{\mu\al}h_{\al\nu})$. We apply  the formula~(\ref{e:U1}) to~(\ref{e:U4}) for $U_j(g^{-1}f)$. These are given in terms of $g^{-1} =(g^{\mu\nu})$. Using  that for an arbitrary function $F(g^{-1})$ we have
\be
\frac{\dd F}{\dd g_{\mu\nu}} = -g^{\mu\al}g^{\nu\beta}\frac{\dd F}{\dd g^{\al\beta}}\,,
\ee
a direct evaluation of $s_i$ and their first and second derivatives  leads to
\begin{subequations} \label{eq:list}
\begin{align}
\bar{s}_1 &= f_{\mu\nu} \bar{g}^{\mu\nu} \, , \\
s^{\mu\nu}_1 &= -\bar g^{\mu\al}\bar g^{\nu\beta} f_{\al\beta} \, , \\
s^{\mu\nu\al\beta}_1 &= \frac{1}{8}f_{\rho\si} \left\{ \left[\bar g^{\mu\rho}\bar g^{\nu\beta}\bar g^{\si\al} + (\mu \leftrightarrow \nu) + (\al \leftrightarrow \beta)  +  (\mu \leftrightarrow \nu) (\al \leftrightarrow \beta) \right]  + \left[ \cdots \right]\big( (\mu,\nu) \leftrightarrow (\al,\beta)\big)\right\} \, , \\ 
&\equiv    {\rm sym}\left\{f_{\rho\si}\bar g^{\mu\rho}\bar g^{\nu\al}\bar g^{\beta\si} \right\}\nonumber \\   &  \nonumber \\
\bar{s}_2 &= \frac{1}{2}f_{\al\beta} f_{\mu\nu} \left(
   \bar{g}^{\al\beta}\bar g^{\mu\nu}- \bar{g}^{\mu \al}\bar g^{\beta
   \nu }\right) \, , \\  
s^{\mu\nu}_2 &=   f_{\rho\si}f_{\la\eta}\left( 
\bar g^{\mu\rho}\bar g^{\nu\la}\bar g^{\si\eta} -\bar g^{\mu\rho}\bar g^{\nu\si}
    \bar g^{\la\eta}    \right)  \, , \\
s^{\mu\nu\al\beta}_2 &= {\rm sym}\left\{f_{\rho\si}f_{\la\eta}\left(\bar g^{\mu\al}\bar g^{\nu\rho}\bar g^{\beta\si}\bar g^{\la\eta} +\frac{1}{2}\bar g^{\mu\rho}\bar g^{\nu\si}\bar g^{\al\la}\bar g^{\beta\eta} 
 -\bar g^{\mu\al}\bar g^{\nu\rho}\bar g^{\beta\la}\bar g^{\si\eta}   -\frac{1}{2}\bar g^{\mu\rho}\bar g^{\nu\la}\bar g^{\al\si}g^{\beta\eta}  \right) \right\}  \, , \\  &  \nonumber \\ 
 \bar{s}_3 &= \frac{1}{6} f_{\al\beta} f_{\rho \nu } f_{\sigma \mu }  \left(2
\bar{g}^{\mu \al}\bar g^{\beta \rho}\bar g^{\nu \sigma }+\bar{g}^{\al\beta}\bar g^{\nu \rho}\bar g^{\mu \sigma }
-3  \bar{g}^{\al\beta}\bar g^{\mu \rho} \bar g^{\nu \sigma } \right) \, , \\
   s^{\mu\nu}_3 &= 
  f_{\sigma \eta }  f_{\rho \la } f_{\al\beta} \left(\bar g^{\mu\si }\bar g^{\nu\rho}
  \bar g^{\eta\la} \bar{g}^{\al\beta}
   - \bar g^{\mu \sigma }\bar g^{\nu\rho} \bar{g}^{\eta \al}\bar g^{\la\beta} +\frac{1}{2} \bar g^{\mu\si}\bar g^{\nu\eta} \bar{g}^{\rho\al}\bar g^{\la \beta } 
   - \frac{1}{2} \bar g^{\mu\si}\bar g^{\nu \eta } \bar g^{\rho\la}\bar{g}^{\al\beta}    \right)  \, , \\
s^{\mu\nu\al\beta}_3 &=   {\rm sym}\left\{f_{\ga\ep} f_{\rho \la } f_{\sigma \eta } \left[ 
\bar g^{\mu\ga}\bar g^{\nu\rho} \bar{g}^{\al\ep}\bar g^{\beta\si}\bar g^{\la\eta} 
+ \bar g^{\mu\ga}\bar g^{\nu\al} \bar{g}^{\beta\rho}\bar g^{\ep\si}\bar g^{\la\eta} 
 +\frac{1}{2}\bar g^{\mu\al}\bar g^{\nu\ga} \bar{g}^{\beta\ep} \bar g^{\rho\la} \bar g^{\si\eta}
 +\frac{1}{2}\bar g^{\mu\ga}\bar g^{\nu\ep} \bar{g}^{\al\rho}\bar g^{\beta\la}\bar g^{\si\eta} 
 \right.\right. \nonumber \\   &  \hspace{2.1cm}  \left.\left.
-\bar g^{\mu\ga}\bar g^{\nu\al} \bar{g}^{\beta\rho}\bar g^{\ep\la} \bar g^{\si\eta}
 -\bar g^{\mu\ga}\bar g^{\nu\rho} \bar{g}^{\al\si}\bar g^{\beta\eta}\bar g^{\ep\la} 
-\frac{1}{2}\bar g^{\mu\ga}\bar g^{\nu\rho} \bar{g}^{\al\ep}\bar g^{\beta\la}\bar g^{\si\eta} 
-\frac{1}{2}\bar g^{\mu\al}\bar g^{\nu\ga} \bar{g}^{\beta\ep}\bar g^{\la\si} \bar g^{\eta\rho}  \right] \right\} \, , \\
    &  \nonumber \\
\bar{s}_4 &= \det\left(\bar g^{-1}f \right) \,, \\
s^{\mu\nu}_4 &=  -\bar{s}_4 \bar g^{\mu\nu}  \, \\
s^{\mu\nu\al\beta}_4 &= \frac{\bar{s}_4}{2}\left(\bar g^{\mu\nu}\bar g^{\al\beta} + \frac{1}{2}\bar g^{\mu\al}\bar g^{\nu\beta}+ \frac{1}{2}\bar g^{\nu\al}\bar g^{\mu\beta}\right) 
\end{align}
\end{subequations}
The operator ${\rm sym}\{\cdots \}$ indicates symmetrization in $ (\mu \leftrightarrow \nu)$, $(\al \leftrightarrow \beta)$ and  $(\mu, \nu)  \leftrightarrow (\al, \beta)$.
 
These are the expressions for the derivatives of the $s_i$ which have to be inserted in the formulas for the variations of de $t_j$ which in turn enter in the expression for 
$\MM^{\mu\nu\al\beta}$. Not surprisingly, the expressions for the variations of $s_2$ and $s_3$ are quite cumbersome. We did not find any further significant simplifications for them in the general case.
\vspace*{0.1cm}

\section{The computation of $\bar{\MM}^{\mu\nu}$}
\label{a:compMbar}

In this Appendix we present more details about the computation of the mass term $\bar{\MM}^{\mu\nu}\equiv\MM^{\mu\nu}(f,\bar g)$ defined in Eq.~(\ref{e:Mmunu}) and used in Eq.~(\ref{e:back_eom}). We have
\be
\bar{\MM}^{\mu\nu} \equiv \biggl\{\frac{1}{\sqrt{-\mathrm{det}g}}\frac{\delta(\sqrt{-\mathrm{det}g}U(g,f))}{\delta g_{\mu\nu}}\biggr\}\bigg|_{g=\bar g} 
    = \frac{1}{2}U(f,\bar{g})\bar{g}^{\mu\nu}-2m^{2}(a_{1} t_1^{\mu\nu}+a_{2} t_2^{\mu\nu}+a_{3}t_3^{\mu\nu}),
\ee
where we have used $\frac{\delta\sqrt{-\mathrm{det}g}}{\delta g_{\mu\nu}}=\frac{1}{2}\sqrt{-\mathrm{det}g}g^{\mu\nu}$ and, from Eq.~(\ref{e:pot4}),
\be
\frac{\delta U(f,g)}{\delta g_{\mu\nu}}   \bigg|_{g=\bar g} = -2m^{2}\biggl(\frac{\delta(a_{1}t_{1}+a_{2}t_{2}+a_{3}t_{3})}{\delta g_{\mu\nu}}\biggr) \bigg|_{g=\bar g} 
 = -2m^{2}(a_{1} t_{1}^{\mu\nu}+a_{2} t_{2}^{\mu\nu}+a_{3} t_{3}^{\mu\nu}).
\ee

The quantities $\ t_j^{\mu\nu}$ can be written in terms of  $ s_{j}^{\mu\nu}$, see Eqs.~(\ref{e:deltaBmunu}) and (\ref{e:deltatmunu}) which are given in Eq.~(\ref{eq:list}).

\section{The equations of motion for the cosmological perturbations}
\label{a:eom}

Here we present the derivation of the equations of motion for the perturbations at first order based on the second-order perturbed part of the action (\ref{e:lag}). In complete generality, these equations of motion have the form

\be \label{eq:1st_order_pert_eom}
\delta {G^{\mu}}_{\nu}+\delta{\MM^{\mu}}_{\nu}=8\pi G \delta {T^{\mu}}_{\nu},
\ee
where $\delta {G^{\mu}}_{\nu}$, $\delta{\MM^{\mu}}_{\nu}$ and $\delta {T^{\mu}}_{\nu}$ stand for the first-order perturbation of the usual Einstein tensor ${G^{\mu}}_{\nu}$, the first-order perturbation of the mass term and the first-order perturbation of the energy-momentum tensor, respectively. The perturbations $\delta {G^{\mu}}_{\nu}$ and $\delta {T^{\mu}}_{\nu}$ can be found in the literature (see e.g.~\cite{Mukhanov:1990me,Kodama:1985bj,Durrer:1994zza}). For the mass term we have

\be \label{eq:deltaGmass}
\delta{\MM^{\mu}}_{\nu}=\delta(\MM^{\mu\rho}g_{\rho\nu})=\delta\MM^{\mu\rho}\bar{g}_{\rho\nu}+\bar{\MM}^{\mu\rho}\delta g_{\rho\nu},
\ee
where
\be
\delta\MM^{\mu\rho}=\frac{\delta}{\delta h_{\mu\rho}}(\MM^{\tau\sigma\alpha\beta}h_{\tau\sigma}h_{\alpha\beta})=2\MM^{\mu\rho\alpha\beta}h_{\alpha\beta}.
\ee 
 $\bar{\MM}^{\mu\rho}$ has already been calculated in Appendix~\ref{a:compMbar}.
We choose the background metric $\bar g_{\mu\nu}$  given by Eq.~(\ref{e:g_phys}), while the  metric $f_{\mu\nu}$ is given by Eq.~(\ref{e:f_phys}) so that we can use Eq.~(\ref{e:MM_phys}) for the components of the mass tensor $\MM^{\mu\nu\alpha\beta}$. We are interested in scalar  perturbations of the metric $g_{\mu\nu}$ which we decompose into Fourier components that evolve independently. Note that we cannot fix a particular gauge since the mass term in the action is not gauge invariant a priori (see, however, the discussion about the "hidden symmetry" for perturbations on Minkowski or de Sitter spacetime in Ref.~\cite{Jaccard:2012ut}). Gauge invariance can be restored by means of the St\"uckelberg trick~\cite{Siegel:1993sk,deRham:2014zqa}, but we are not doing this here. The metric perturbation of a Fourier component is
\be
h_{\mu\nu}(\tau,k) \equiv \delta g_{\mu\nu}=
\begin{pmatrix}
-2\phi & iak_{j}B \\
iak_{i}B & 2a^{2}(\psi\delta_{ij}-k_{i}k_{j}E)
\end{pmatrix}.
\ee
The energy-momentum tensor up to first order in scalar perturbations is given by
\be
{T^{\mu}}_{\nu}=
\begin{pmatrix}
-\bar{\rho}-\delta\rho & -a(\bar{\rho}+\bar{p})(ik_{j}v-ik_{j}B)  \\
a^{-1}(\bar{\rho}+\bar{p})ik^{i}v & (\bar{p}+\delta p)\delta^{i}_{j}
\end{pmatrix}.
\ee
The first-order perturbation equation, 
$\delta {G^{0}}_{0}+\delta{\MM^{0}}_{0}=M_P^{-2} \delta {T^{0}}_{0}$,
then becomes
\be\label{ea:00}
\begin{split}
 & \biggl\{\frac{2 k^2}{a^2}+ 3m^2 \biggl( 2 c_3  (r-2) (r-1)+ c_4 (r-1)^2+
   r^2-6 r+6\biggr)\biggr\}\psi +\frac{2 H k^2}{a} B \\
   &- m^2 \biggl\{2 c_3  (r-2)
   (r-1)+c_4  (r-1)^2+ r^2-6r+6\biggr\} k^2 \text{E}\\
   &- \biggl\{6 H^2 + m^2(r-1) \biggl(c_4
   (r-1)^2+c_3 (r-4) (r-1)-3r+6\biggr)\biggr\} \phi -2 H k^2
   \dot{\text{E}} +6 H \dot{\psi}=M_P^{-2}  \delta \rho.
\end{split}
\ee
Equation
$ 
\delta {G^{0}}_{i}+\delta{\MM^{0}}_{i}=M_P^{-2} \delta {T^{0}}_{i} 
$
is
\be\label{ea:0inew}
m^2 \frac{(r-1) r^2 \bigl(c_3
   (r-3)+c_4 (r-1)\bigr)+ (3-2 r) r^2}{r+1}B+\frac{2H}{a} \phi -\frac{2}{a}  \dot\psi 
=M_P^{-2} \left(\bar{p}+\bar{\rho }\right)(v-B).
\ee
Equation
$
\delta {G^{i}}_{i}+\delta{\MM^{i}}_{i}=M_P^{-2} \delta {T^{i}}_{i}
$
reads
\be\label{ea:ii}
\begin{split}
 &  m^2\biggl\{ c_3 (r-3) (r-1)+ r^2-8
   r+9\biggr\}k^2 \text{E} \\
  &+ \biggl\{- 3m^2 \biggl(2 c_3 (r-2) (r-1)+ c_4
   (r-1)^2+r^2-6 r+6\biggr)+12 \dot{H}+18 H^2-2 \frac{k^2}{a^2}\biggr\}  \phi\\
   &-
   \bigg\{3m^2 \biggl( c_3 (r-3) (r-1)+ r^2-8 r+9\biggr)+2
   \frac{k^2}{a^2}\biggr\} \psi  \\
   &+ 2 k^2 \ddot{\text{E}}-18
   H \dot{\psi}+6 H \dot{\phi}-6 \ddot{\psi}+6 H k^2 \dot{\text{E}}-\frac{2 k^2 }{a}\dot{B}-\frac{4
    H k^2}{a}B=3 M_P^{-2} \text{$\delta $p}.
\end{split}   
\ee

Finally, the longitudinal, traceless part of the ($ij$) component of the equation of motion,
\be \label{eq:longit_traceless}
\biggl(\hat{k}_{i}\hat{k}^{j}-\frac{1}{3}\delta^{j}_{i}\biggr)\left(\delta {G^{i}}_{j}+\delta{\MM^{i}}_{j}\right)=M_P^{-2} \biggl(\hat{k}_{i}\hat{k}^{j}-\frac{1}{3}\delta^{j}_{i}\biggr)\delta {T^{i}}_{j}
\ee 
(where $\hat{k}_{i}$ is the unit wave vector), reads
\be
 m^2 \biggl\{ c_3 (r-1) r+ r^2-2 r\biggr\}   \text{E} +\frac{\dot{B}}{a}+\frac{\psi}{a^2}- \ddot{\text{E}}-3 H \dot{\text{E}}+\frac{ \phi
   }{a^2}+\frac{2 H}{a}B =0.
\ee
For the static situation, $H=\dot H=0$ and vanishing matter perturbations, this system reduces to (\ref{e:osc}).

~ 
\end{widetext}

\bibliography{massiveG}
\end{document}